# Leveraging Public Cloud Infrastructure for Real-time Connected Vehicle Speed Advisory at a Signalized Corridor

Hsien-Wen Deng[1], M Sabbir Salek, Ph.D.[2*], Mizanur Rahman, Ph.D.[3], Mashrur Chowdhury, Ph.D.[2], Mitch Shue[1], and Amy W. Apon, Ph.D.[1]

*[1]School of Computing, Clemson University, Clemson, SC 29631, USA*
*[2]Glenn Department of Civil Engineering, Clemson University, Clemson, SC 29631, USA*
*[3]Department of Civil, Construction & Environmental Engineering, The University of Alabama, Tuscaloosa, AL 35487, USA*
*[*]Corresponding Author (E-mail: msalek@clemson.edu)*

**ABSTRACT**

In this study, we developed a real-time connected vehicle (CV) speed advisory application that uses public cloud services and tested it on a simulated signalized corridor for different roadway traffic conditions. First, we developed a scalable serverless cloud computing architecture leveraging public cloud services offered by Amazon Web Services (AWS) to support the requirements of a real-time CV application. Second, we developed an optimization-based real-time CV speed advisory algorithm by taking a modular design approach, which makes the application automatically scalable and deployable in the cloud using the serverless architecture. Third, we developed a cloud-in-the-loop simulation testbed using AWS and an open-source microscopic roadway traffic simulator called Simulation of Urban Mobility (SUMO). Our analyses based on different roadway traffic conditions showed that the serverless CV speed advisory application meets the latency requirement of real-time CV mobility applications. Besides, our serverless CV speed advisory application reduced the average stopped delay (by 77%) and the aggregated risk of collision (by 21%) at signalized intersection of a corridor. These prove the feasibility as well as the efficacy of utilizing public cloud infrastructure to implement real-time roadway traffic management applications in a CV environment.

## 1. INTRODUCTION

### 1. 1. Background and Motivation

In transportation cyber-physical systems (TCPS), the interaction between cyber and physical systems makes it possible to develop CV-based real-time roadway management applications (Deka and Chowdhury, 2018). However, to develop a real-time feedback-based interaction between cyber systems and physical systems, high-performance computing infrastructure is required to process the heterogeneous data from different sources. The departments of transportation (DOTs) would have to rely on traditional server infrastructure installed in the traffic management centers (TMCs) to implement such applications at present condition. This requires significant human resources and investments for installation, configuration, operation, maintenance, and upgrade. While edge or fog computing offers a viable solution to deploy CV-based real-time roadway traffic management applications in a TCPS environment (Omoniwa et al., 2019), there are also issues related to edge computing-based deployments, such as wireless communication range (Xu et al., 2017), let alone the requirement of significant financial investments. The recent evolution of public cloud infrastructure has made it possible to support real-time CV-based TCPS applications in the cloud (Deng et al., 2021). As a result, many public and private transportation agencies are nowadays considering public cloud infrastructure for replacing their on-premise roadway traffic management operation. For example, Elizabeth River Crossing, a private construction company, in partnership with Virginia Department of Transportation saved over $200,000 in infrastructure maintenance cost utilizing cloud services provided by Amazon Web Services or AWS (i.e., a popular public cloud service provider) ("The Cloud Helps Transportation Agency Keep Region Moving Despite Disruption," 2020). Additionally, most commercial cloud service providers now offer serverless solutions, for example, Lambda ("What is AWS Lambda? - AWS Lambda," 2022) offered by AWS, Azure Functions ("Azure Functions Serverless Compute | Microsoft Azure," 2022) offered by Microsoft

Azure, that remove the burden of establishing server instances and enable developers to focus primarily on application development, such as CV and Internet of Things (IoT) applications.

Cloud infrastructure can be utilized to develop applications in a server-based manner or a serverless manner. In a server-based cloud application, the application developers are required to establish server instances, e.g., Amazon Elastic Compute Cloud or Amazon EC2 ("What is Amazon EC2? - Amazon Elastic Compute Cloud," 2022), and configure coding platforms in the cloud that will support the application. For instance, Ning et al. (Ning et al., 2019) utilized a cloud-based fog computing architecture to implement real-time roadway traffic management. Jin et. al. (Jin et al., 2020) presented a method of constructing cloud-based mobility services for connected and automated vehicle highway systems. These studies used a traditional server-based architecture to develop real-time CV-based roadway traffic management applications. On the other hand, in a serverless cloud-based application, the application developers do not need to establish the server instances as the computational resources are managed by the cloud itself based on the computing requirement of an application. Therefore, if the application requires to utilize additional or less capacity in the database or computing infrastructure due to an increase or decrease in the number of CVs subscribed to the application, the cloud service provider will automatically employ those additional resources (i.e., scale up) or reduce them (i.e., scale down) to accommodate this change. This makes a serverless cloud-based CV application resource-efficient, scalable, and automated. Also, since the cloud service provider oversees configuring and managing the underlying infrastructure, the application developers can solely focus on application development while using a serverless architecture, which makes serverless cloud an attractive option to develop real-time CV-based roadway traffic management applications. However, deploying a real-time public cloud-based roadway traffic management application for CVs in a serverless manner requires developing a feasible serverless cloud architecture utilizing the available public cloud services as well as developing an algorithm for the CV application that is deployable through the serverless cloud architecture while meeting the latency requirements of a real-time CV application. Recently, Deng et al. (Deng et al., 2021) utilized AWS serverless infrastructure to develop and field-validate a traffic surveillance application to compute the average speed of CVs in a TCPS environment, where CVs sent their speed information directly to a database in AWS. In this study, we used a serverless architecture based on public cloud services for developing a real-time CV-based roadway traffic management application that requires the cloud infrastructure to perform computation using data coming from not only CVs but also transportation infrastructure (such as traffic signals at an intersection) in real-time while meeting the latency requirement of CV mobility applications.

In this study, we developed a CV speed advisory application, i.e., an application that provides each CV with an advised speed that changes dynamically based on various factors, such as the CV's location, surrounding traffic condition, and signal phase and timing of the traffic signal at an intersection that the CV is approaching. Our CV speed advisory application leverages existing public cloud infrastructure in a serverless manner with a goal to minimize the stopped delay experienced by CVs while passing through a signalized corridor. In this TCPS environment, serverless public cloud infrastructure (as cyber systems) interacts with CVs and connected traffic signals (both as parts of physical systems), as shown in Fig. 1. In general, serverless public cloud infrastructure offers three types of services: (i) Function as a Service (FaaS), where a roadway traffic management application, such as a CV speed advisory application, can run, (ii) Platform as a Service (PaaS), where computing, data streaming, and database management services operate, and (iii) Infrastructure as a Service (IaaS), which is managed by the public cloud service providers.

The serverless public cloud infrastructure features a pay-as-you-go model without having to manage the underlying computing infrastructure. As shown in Fig. 1, the serverless public cloud-based CV speed advisory application requires real-time CV trajectory information, such as the location and speed of the CVs, which can be sent to the cloud from the CVs in the form of basic safety messages (BSMs). On the other hand, the CVs will receive speed advisories from the cloud, which the CVs can download directly from a database residing in the cloud. As shown in Fig. 1, the CV speed advisory application would also require other information, such as real-time signal phase and timing (SPaT) information, which the application can collect from the connected traffic signals, and the location and attributes of the roadway traffic signs (e.g., regulatory and warning signs), which the application can collect from transportation asset management database.

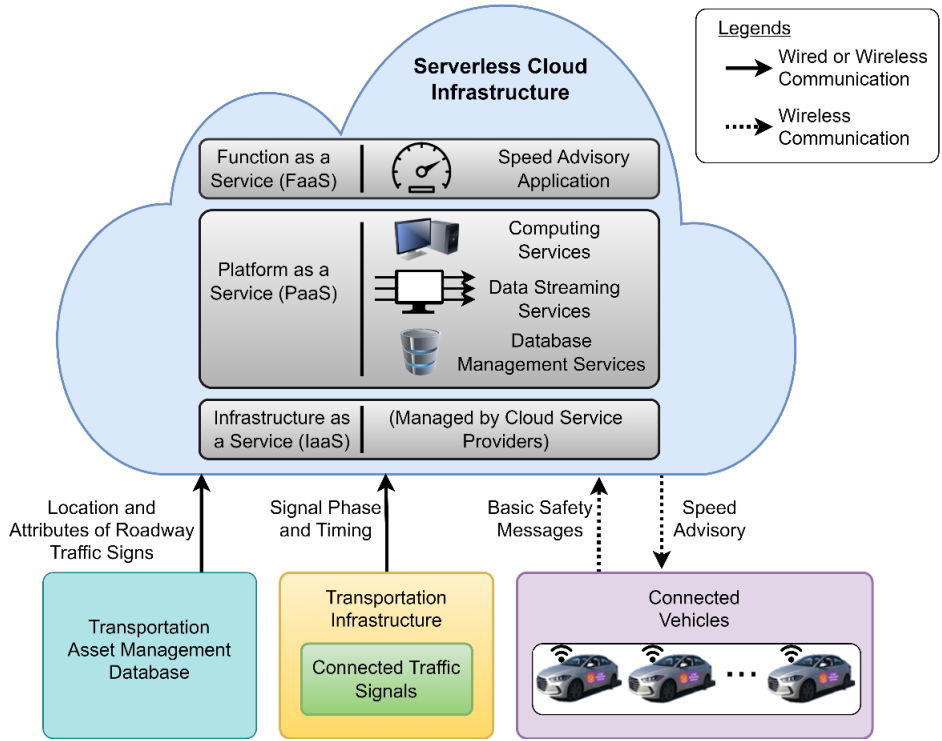

**Fig. 1.** Serverless public cloud architecture for a CV speed advisory application in a TCPS environment.

## 1. 2. Contribution

This study developed a real-time CV speed advisory application leveraging the existing public cloud infrastructure for real-time CV-based roadway traffic management, which, we believe, is a promising resource and cost-efficient alternative to deploying such applications through traditional server or roadside infrastructure. The primary contributions of this study are (i) to develop a serverless public cloud architecture (using AWS) for a real-time CV speed advisory application in a TCPS environment, and (ii) to develop an optimization-based real-time CV speed advisory algorithm using a modular design approach, which makes the application scalable and deployable using the developed serverless public cloud architecture in terms of end-to-end latency requirement of CV mobility applications.

## 2. LITERATURE REVIEW

In this section, we present a review of the existing studies related to (i) cloud-based CV applications, and (ii) optimal speed advisory algorithms. Based on the review, we present the research gaps related to cloud-based CV speed advisory application, which is addressed in this study.

### 2. 1. Cloud-based CV Applications

Cloud infrastructures can effectively communicate with CVs and transportation infrastructures through vehicle-to-infrastructure (V2I) and infrastructure-to-infrastructure (I2I) communication, respectively. The communication among cloud infrastructures and transportation infrastructures can be via either wired or wireless communication, whereas, to communicate with CVs, the cloud infrastructures must utilize wireless communication. The wireless communication between CVs and cloud infrastructures leverages connectivity to the cloud over the internet via cellular networks ("Designing Next Generation Vehicle Communication with AWS IoT Core and MQTT: AWS Whitepaper," 2024). Services in the cloud then aggregate and analyze the data from CVs and transportation infrastructure to generate the required information corresponding to the cloud applications. Cloud-based CV applications, such as, vehicular cloud, vehicular ad-hoc network (VANET), edge, and fog computing-based CV applications are widely discussed in the literature. For instance, Ning et al. utilized a cloud-based fog computing architecture to implement real-time roadway traffic management strategy (Ning et al., 2019). Li et. al. provided a maximum value density-based heuristic algorithm through vehicular edge cloud computing to achieve energy usage efficiency for roadway traffic (Li et al., 2020). Jin et. al. presented a method of constructing cloud-based mobility services for highway applications to accommodate connected and automated vehicles (Jin et al., 2020).

Several public cloud service providers now feature reference architectures for developing cloud-based CV applications using their cloud services, such as AWS connected vehicle reference architecture ("AWS Connected Vehicle Reference Architecture," 2023) and Azure connected vehicle fleet architecture ("Microsoft Connected Fleet Reference Architecture," 2024). Studies that utilize public cloud services for CV application development are also growing. Among them, Khaled introduced an AWS-based vehicular diagnostic application utilizing a digital twin-based algorithm running in the cloud (Khaled, 2021). The author in (Khaled, 2021) evaluated the application using vehicle motion and hardware-related data, such as vehicle speed, engine and motor speed, and battery state of charge data, from 1,000 CVs and reported that the application was effective in detecting and notifying the drivers about potential system failures. Liao et al. developed a cloud-based cooperative CV ramp merging application, in which data from CVs are utilized in the cloud for development and real-time synchronization of digital twins of the CVs and their respective drivers with their real-world counterparts (Liao et al., 2022). The digital twins were then used to determine optimal ramp merging advisories for the CVs. The authors in (Liao et al., 2022) validated their application in real-world using three passenger CVs. In our previous work (Deng et al., 2021), Deng et al. utilized AWS serverless infrastructure to develop a traffic surveillance application to compute the average speed of CVs in a TCPS environment. However, the authors in (Deng et al., 2021) only considered communication between the cloud and the CVs. In this study, we used a serverless architecture in a public cloud for a real-time CV mobility application that requires the cloud infrastructure to perform computation using data coming from both CVs and transportation infrastructure in real-time while meeting the strict latency requirement of the CV mobility applications. To this end, the application developed in this study provides a blueprint for future studies for developing other real-time serverless public cloud-based CV mobility applications.

## 2. 2. Optimal Speed Advisory Algorithms

Optimal speed advisory algorithms help CVs navigate through a signalized corridor efficiently in terms of reduced stopped delay, fuel consumption, and $CO_2$ emission, and they have been studied extensively in the literature. Many studies referred to this type of algorithm as the Green Light Optimal Speed Advisory (GLOSA) algorithm (Bradaï et al., 2016; Stebbins et al., 2016, 2017; Suzuki and Marumo, 2018; Pariota et al., 2019; Zhang et al., 2020). For instance, Suzuki and Marumo developed a GLOSA system that projects a green rectangle on the roadway through the head-up display of a GLOSA-enabled vehicle (Suzuki and Marumo, 2018). Stebbins et al. combined model predictive control (MPC) with state-space reduction and GLOSA to yield efficient trajectories for the CVs (Stebbins et al., 2016). However, few studies considered platoon formation in GLOSA. Among them, Stebbins et al. developed a platoon-based optimization technique for GLOSA (Stebbins et al., 2017). The authors included a safety constraint in their optimization model considering that the human drivers may not follow an advised speed if they feel that they will not be able to stop if needed while approaching an intersection. Zhao et al. developed a platoon-based MPC to optimize fuel consumption which enables a platoon of vehicles to pass an intersection within a traffic signal system's green interval (Zhao et al., 2018), where the model's efficacy was evaluated for different CV penetration rates. However, none of these studies considered a real-time implementation of the platoon-based GLOSA system for speed advisories in a signalized corridor that is "deployable" in a commercial cloud-based TCPS environment. In this study, we developed a platoon-based real-time CV speed advisory application following a modular design approach for the underlying optimization algorithms, which makes it deployable using serverless public cloud services.

## 3. CV SPEED ADVISORY APPLICATION ARCHITECTURE AND ALGORITHMS

In this section, we present the architectural and algorithmic details of our serverless CV speed advisory application that we developed using public cloud infrastructure.

## 3. 1. Public Cloud-based Serverless Architecture

AWS maintains a vast cloud infrastructure and services catalog, which makes it secure, scalable, and available for developing real-time CV-based roadway traffic management applications (Deng et al., 2021). Besides, AWS offers various serverless services, such as AWS Lambda ("What is AWS Lambda? - AWS Lambda," 2022) as a FaaS, and DynamoDB ("What Is Amazon DynamoDB? - Amazon DynamoDB," 2022) as a PaaS, that can be used to develop applications without being concerned about establishing or maintaining any server instances. Such serverless services generally follow pay-as-you-go billing models that make the serverless architectures cost-effective as we mentioned before ("Serverless Computing – Amazon Web Services," 2022). Thus, in this study, we developed a serverless cloud-based real-time roadway traffic management application utilizing the serverless services offered by AWS, such as AWS Lambda. In Fig. 2, we present a serverless cloud architecture showing the computing resources, databases, and streaming services integrated to support a real-time speed advisory application for CVs using AWS. The serverless architecture eliminates the need for developers to manage traditional server infrastructure. Thus, we only need to focus on developing the application using relevant AWS services.

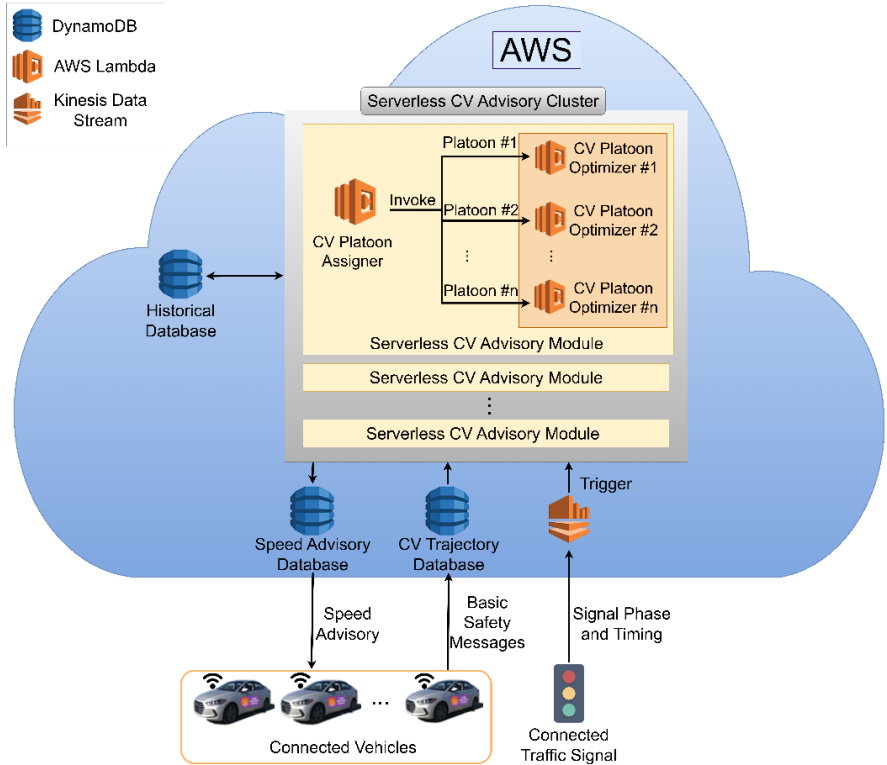

**Fig. 2.** Details of the serverless architecture using AWS services for a CV speed advisory application.

The serverless architecture (shown in Fig. 2) employs the following AWS services: 1) DynamoDB as a database service, 2) Kinesis Data Stream (KDS) as a real-time data streaming service, and 3) AWS Lambda as a compute service. We used DynamoDB, i.e., a NoSQL database service with a key-value structure ("What Is Amazon DynamoDB? - Amazon DynamoDB," 2022), for creating our databases. We created a CV trajectory database to update the CVs' trajectory information, and a speed advisory database to store speed advisory results from which the CVs can download their corresponding speed advisories in real-time. For each traffic signal, we created a historical database to save and update the distances between CVs and the traffic signal in real-time. We utilized Kinesis Data Stream (KDS), a real-time data streaming service ("Amazon Kinesis Data Streams - Data Streaming Service - Amazon Web Services," 2022) in AWS, to send a message from each traffic signal to the cloud every second to trigger (i.e., launch the target program automatically) the serverless functions in CV advisory cluster. AWS Lambda ("What is AWS Lambda? - AWS Lambda," 2022) is the serverless compute service at the core of this serverless architecture. We designed a group of AWS Lambda functions to form a serverless CV advisory cluster that gets triggered by KDS for each traffic signal. Each cluster contains multiple serverless CV advisory modules that process information from the CVs. To meet the latency requirement of a real-time CV mobility application, i.e., less than or equal to 1000 ms (Fehr, 2014; Islam et al., 2020), we defined the capacity of each serverless CV advisory module in terms of the maximum number of CVs to be processed, which was 50 CVs per module in our AWS implementation, and ran all the CV advisory modules in parallel. The usage of parallel computing in a cluster makes our CV speed advisory application fast and scalable. Note that, the serverless architecture presented in Fig. 2 should be considered as an example of utilization of serverless public cloud infrastructure for real-time roadway traffic management. Indeed, similar serverless public cloud-based architectures can be developed for any real-time roadway traffic management applications by assigning an appropriate serverless database service to store the data to processed or already processed, an appropriate compute service to run any code for data processing or analyses, a real-time data streaming service to periodically trigger the application, and arranging them in a fashion that enables utilization of parallel computing to make the application fast and scalable on-demand.

In the serverless CV speed advisory architecture presented in Fig. 2, there are two types of programs in each serverless CV advisory module: 1) a CV platoon assigner, and 2) a set of CV platoon optimizers. A CV platoon assigner is an AWS Lambda function that has the necessary information related to its corresponding traffic signal and intersection, such as the physical location, signal phase duration of the traffic signal, and the posted speed limit on the roadway approaching that intersection. Once the cluster is triggered, the CV platoon assigner performs the following tasks: 1) collect information from both traffic signals and CVs, 2) split the CVs into platoons based on the CVs' gap information (based on the method discussed in the first subsection titled 'CV Platoon Identification' of the following section), 3) compute a speed advisory for only the leader CV of each platoon (based on the method discussed in the last two subsections of the following section) save the speed advisory for the leader CV of each platoon into the speed advisory database. Then, for each platoon, the CV platoon assigner invokes a set of CV platoon optimizers based on the number of platoons identified. A CV platoon optimizer is also a serverless process, i.e., an AWS Lambda, that is responsible for its corresponding CV platoon. It computes speed advisories for the follower CVs in that platoon to help them pass the intersection while maintaining the minimum safety distances and operating within the roadway speed limit. The results, i.e., the speed advisories for the follower CVs, generated from the CV platoon optimizers are then stored in the speed advisory database.

In the real world, each CV generates BSMs and each traffic signal generates SPaT messages. In our serverless CV speed advisory application, each CV uploads a filtered BSM including the CV's ID, location, speed, and the gap with its immediate leading CV into the CV trajectory database. Each traffic signal sends a filtered SPaT message every second containing the current traffic signal phase and the remaining time of that phase through KDS. Our optimization-based speed advisory algorithm deployed in each serverless CV advisory cluster utilizes these BSMs and signal phase and timing messages to generate speed advisories for the CVs in real-time.

### 3. 2. Algorithms

We utilized an optimization-based CV speed advisory algorithm that we deployed in AWS to minimize the stopped delay for CVs at signalized intersections. We adopted a modular design approach for the speed advisory algorithm so that the application can automatically and efficiently scale up or down depending on the application requirement. To this end, our CV speed advisory algorithm is divided into three parts to leverage the serverless cloud services offered by AWS: 1) CV platoon identification (presented in subsection 3.2.1), 2) optimization-based speed advisory algorithm for the leader CVs of the platoons (presented in subsection 3.2.2), and 3) optimization-based speed advisory algorithm for the follower CVs of the platoons (presented in subsection 3.2.3). The first two algorithms run in an AWS Lambda, i.e., a CV platoon assigner, whereas the last algorithm utilizes a set of AWS Lambdas, i.e., a set of CV platoon optimizers in parallel (see Fig. 2). For the last algorithm, the number of AWS Lambdas that should be deployed in parallel is selected automatically based the number of CV platoons. Thus, the utilization of parallel computing for the last algorithm makes the application scalable while meeting the strict latency requirement for CV mobility applications. In this section, we present the algorithms in detail.

*3.2.1 CV Platoon Identification*

We form CV platoons based on whether they can pass a signalized intersection within the available time of the current green time or the next green time, i.e., $t_{avail}(k)$, measured at the $k^{th}$ timestamp (Fig. 3 and Table 1 present symbols used in developing the CV speed advisory algorithm). Therefore, to be identified as a platoon of $N$ number of CVs, the last or $N^{th}$ CV of the platoon must be able to pass the intersection within the available time, i.e., meet the following criterion:

$$\min t_{N,int}(k) \leq t_{avail}(k) \quad (1)$$

where, $t_{N,int}(k)$ denotes the estimated time taken by the $N^{th}$ CV to reach the target intersection from its location at the $k^{th}$ timestamp, and $t_{avail}(k)$ is the available time to pass the target intersection calculated at the $k^{th}$ timestamp. To estimate the minimum of $t_{N,int}(k)$, we consider the total time required by the $N^{th}$ CV to accelerate from its current speed ($S_N(k)$) to the maximum speed based on the roadway speed limit ($S_{max}$) using its maximum acceleration ($a_{Acc}$) and then continue to travel at $S_{max}$ until it reaches the intersection, which is given by the following equation (according to Newton's equations of motion),

$$\min t_{N,int}(k) = \frac{S_{max}-S_N(k)}{a_{Acc}} + \frac{1}{S_{max}}\left[d_{N,int}(k) - \frac{(S_{max})^2-S_N^2(k)}{2a_{Acc}}\right] \quad (2)$$

where, $d_{N,int}(k)$ is the distance from the $N^{th}$ follower CV in a platoon to the target intersection at the $k^{th}$ timestamp. First part of the right side of the above equation gives the minimum time required by the $N^{th}$ CV to accelerate from $S_N(k)$ to $S_{max}$, and the second part of the equation gives the time required by the $N^{th}$ CV to reach the intersection at a constant speed (i.e., $S_{max}$) after it achieves $S_{max}$. Thus, (2) estimates the minimum time required by the $N^{th}$ CV of the platoon to reach the intersection. We will explain how the time spent in constant speed is obtained for the leader CV of a platoon in the next subsection.

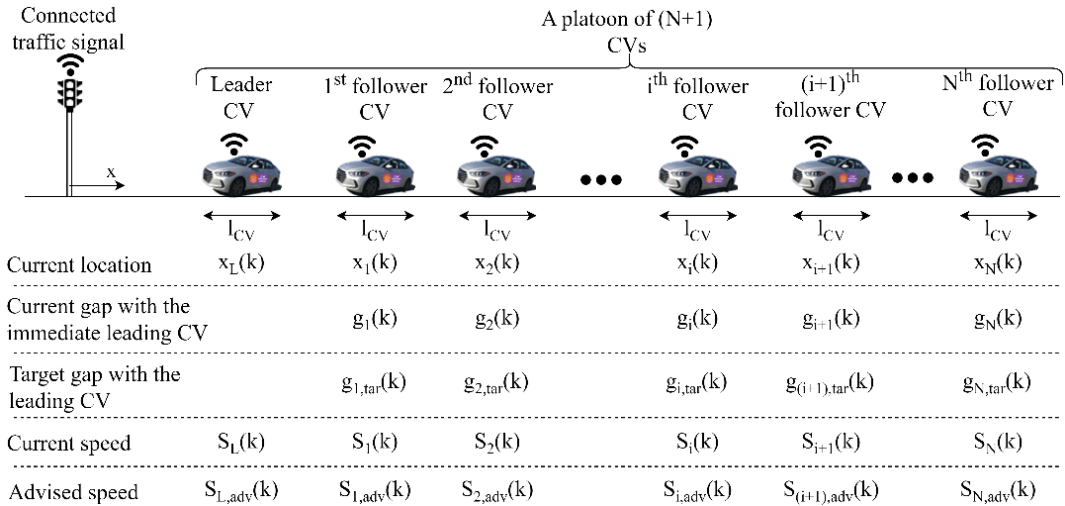

**Fig. 3.** Symbols used in the CV speed advisory algorithm.

**Table 1**
Symbols used in the CV speed advisory algorithm.

| Symbol | Meaning |
|---|---|
| $k$ | $k$ in parenthesis following a symbol indicates value at the $k^{th}$ timestamp |
| $g_{i,tar}$ | Target gap of the $i^{th}$ follower CV |
| $g_{stand}$ | Constant standstill gap |
| $T_g$ | Constant time gap |
| $S_{max}$ | Maximum speed, which is same as the roadway speed limit |
| $a_{Acc}$ | Maximum acceleration |
| $a_{Brk}$ | Maximum braking deceleration |
| $a_{const}$ | Constant acceleration; $a_{const} = a_{Acc}$ if the CV is accelerating, or $a_{Brk}$ if decelerating |
| $delay_L$ | Additional estimated delay experienced by the leader CV of a platoon while following $S_{L,adv}$ compared to following $S_{max}$ |
| $d_{i,int}$ | Distance from the $i^{th}$ follower CV in a platoon to the target intersection |
| $d_{L,int}$ | Distance from the leader CV of a platoon to the target intersection |
| $d_{L,constAcc}$ | Estimated distance covered by the leader CV of a platoon while accelerating from $S_L$ to achieve a target speed ($S_{L,tar}$) |
| $d_{L,constSpd}$ | Estimated distance covered by the leader CV of a platoon while operating at a target speed ($S_{L,tar}$) from the moment it achieves $S_{L,tar}$ |
| $t_{N,int}$ | Estimated total time required by the $N^{th}$ follower CV of a platoon to reach the intersection from its location ($x_N$) |
| $t_{L,constAcc}$ | Estimated time required by the leader CV to accelerate from $S_L$ to a target speed ($S_{L,tar}$) |
| $t_{L,constSpd}$ | Time required by the leader CV of a platoon to reach the intersection while operating at a target speed ($S_{L,tar}$) from the moment it achieves $S_{L,tar}$ |
| $t_{remain}$ | Remaining time of the current green interval |
| $t_{avail}$ | Available time to pass an intersection |
| $t_G$ | (Minimum) green interval |
| $t_{AR}$ | All red interval |
| $t_Y$ | Yellow interval |

We assumed 100% CV penetration on the signalized corridor considered in this study. There are two cases to consider based on the current phase of the traffic signal at the target intersection that the CVs are approaching; case I: the platoon can pass the intersection within the current green interval, and case II: the platoon can pass the intersection in the next green interval. For case I, the available time to reach the intersection before the signal turns red is,

$$t_{avail}(k) = t_{remain}(k) \tag{3}$$

where, $t_{remain}(k)$ is the remaining green interval, whereas, for case II, this available time is an aggregate of the remaining green interval and the other intervals till the next green interval, i.e., sum of the minimum green intervals ($\sum t_G$) and yellow intervals ($\sum t_Y$) for the other approaches in the intersection, and sum of the all-red intervals ($\sum t_{AR}$);

$$t_{avail}(k) = t_{remain}(k) + \sum t_G + \sum t_Y + \sum t_{AR} \tag{4}$$

### 3.2.2 Speed Advisory for the Leader CVs of the Platoons

For the leader CV of a platoon, the speed advisory is determined based on whether the platoon is a case I platoon or a case II platoon. For the case I platoons, the speed advisory algorithm attempts to assist the CVs to cross the intersection as fast as possible while operating within the roadway speed limit, $S_{max}$. Therefore, for case I, the leader CVs are simply advised with the roadway speed limit, $S_{max}$, as the speed advisory. For the case II platoons, the speed advisories for the leader CVs are found through an optimization with an objective to reduce the estimated delay to pass the intersection.

For a case II platoon, our objective function of the optimization for determining the advisory speed for the leader CV is the estimated delay experienced by the leader CV while traveling from its current state till it reaches the target intersection. In this context, "delay" (denoted as $delay_L(k)$ calculated from the $k^{th}$ timestamp) is estimated as the additional time required by the leader CV to reach the intersection using the advised speed, $S_{L,adv}$, compared to the lowest possible time to reach the intersection using the maximum speed, i.e., $S_{max}$, which is set to be the same as the speed limit. Thus, the objective function for this optimization is considered as this additional estimated delay for the leader CV, which is given by the following expressions,

$$\min_{S_{L,adv}} delay_L(k) \tag{5}$$

where, 
$$delay_L(k) = \left(t_{L,constAcc}(k) + t_{L,constSpd}(k)\right)_{for\ S_{L,adv}} - \left(t_{L,constAcc}(k) + t_{L,constSpd}(k)\right)_{for\ S_{max}} \tag{6}$$

Here, $\left(t_{L,constAcc}(k) + t_{L,constSpd}(k)\right)_{for\ S_{L,adv}}$ and $\left(t_{L,constAcc}(k) + t_{L,constSpd}(k)\right)_{for\ S_{max}}$ both consist of two periods:

- Acceleration period, $t_{L,constAcc}(k)$: the time required to accelerate from the leader CV's current speed, i.e., $S_L(k)$, to the advised speed, i.e., $S_{L,adv}(k)$ or $S_{max}$; and
- Constant speed period, $t_{L,constSpd}(k)$: the time required to reach the intersection at a constant speed, $S_{L,adv}(k)$ or $S_{max}$, after achieving $S_{L,adv}(k)$ or $S_{max}$.

Here, we only discuss how to estimate the above two periods for $S_{L,adv}(k)$ as the same steps are followed to estimate the two periods for $S_{max}$. The required time to accelerate from $S_L(k)$ to $S_{L,adv}(k)$ is given by,

$$t_{L,constAcc}(k) = \frac{S_{L,adv} - S_L(k)}{a_{const}} \tag{7}$$

where, $a_{const} = a_{Acc}$ if $S_{L,adv}(k) > S_L(k)$, and $a_{const} = a_{Brk}$ if $S_{L,adv}(k) < S_L(k)$. Here, $a_{Brk}$ denotes the maximum braking deceleration. Then, we estimate the distance covered during the acceleration period. Distance covered while accelerating from $S_L(k)$ to $S_{L,adv}(k)$ is given by,

$$d_{L,constAcc}(k) = \frac{\left(S_{L,adv}(k)\right)^2 - S_L^2(k)}{2a_{const}} \tag{8}$$

To determine $t_{L,constSpd}(k)$, first, we need to estimate the distance covered while operating at a constant speed, $S_{L,adv}(k)$, i.e., $d_{L,constSpd}(k)$, which can be obtained by subtracting $d_{L,constAcc}(k)$ from the distance of the leader CV from the target intersection, i.e., $d_{L,int}(k)$,

$$d_{L,constSpd}(k) = d_{L,int}(k) - d_{L,constAcc}(k) = d_{L,int}(k) - \frac{\left(S_{L,adv}(k)\right)^2 - S_L^2(k)}{2a_{const}} \tag{9}$$

Now, we can estimate $t_{L,constSpd}(k)$ for $S_{L,adv}(k)$ as follows,

$$t_{L,constSpd}(k) = \frac{d_{L,constSpd}(k)}{S_{L,adv}(k)} = \frac{1}{S_{L,adv}(k)}\left[d_{L,int}(k) - \frac{\left(S_{L,adv}(k)\right)^2 - S_L^2(k)}{2a_{const}}\right] \quad (10)$$

Similarly, $t_{L,constAcc}(k)$ and $t_{L,constSpd}(k)$ for $S_{max}$ can be written as follows,

$$t_{L,constAcc}(k) = \frac{S_{max} - S_L(k)}{a_{const}} \quad (11)$$

$$t_{L,constSpd}(k) = \frac{d_{L,constSpd}(k)}{S_{max}} = \frac{1}{S_{max}}\left[d_{L,int}(k) - \frac{(S_{max})^2 - S_L^2(k)}{2a_{const}}\right] \quad (12)$$

Therefore, we can now estimate the delay experienced by the leader CV while traveling from its current state until it reaches the target intersection by substituting the terms derived in (7), (10), (11) and (12) into (6),

$$delay_L(k) = \left(d_{L,int}(k) + \frac{S_L^2(k)}{2a_{const}}\right)\left[\frac{1}{S_{L,adv}(k)} - \frac{1}{S_{max}}\right] - \frac{S_{L,adv}(k) - S_{max}}{2a_{const}} \quad (13)$$

For this speed advisory optimization for the case II platoons' leader CV, we consider the following constraint,

$$S_{max} - 10\ mph \leq S_{L,adv} \leq UB \quad (14)$$

where, $UB = \begin{cases} \min\left(S_{max}, \frac{d_{L,int}(k)}{t_{avail}(k)}\right) & if\ \frac{d_{L,int}(k)}{t_{avail}(k)} \geq (S_{max} - 10\ mph) \\ (S_{max} - 10\ mph) & if\ \frac{d_{L,int}(k)}{t_{avail}(k)} < (S_{max} - 10\ mph) \end{cases}$

The above constraint sets the lower and upper bounds to the speed advisory for the case II platoons' leader CVs. The lower bound makes sure that the case II platoons' leader CVs are not advised speeds that are too low compared to the roadway speed limit. To ensure this, the lower bound is set to 10 miles per hour (mph) below the roadway speed limit, $S_{max}$. We chose this threshold to be 10 mph because a threshold less than 10 mph, for example, 5 mph below the speed limit would leave a small window to select the advisory speeds, and a threshold greater than 10 mph, for example, 15 mph below the speed limit, might cause selecting advisory speeds that are too low compared to the roadway speed limit. On the other hand, the upper bound ensures that 1) the advised speeds do not exceed the roadway speed limit, $S_{max}$, and 2) the leader CVs do not arrive at the intersection early before the signal turns green again.

Note that, if $\frac{d_{L,int}(k)}{t_{avail}(k)} < (S_{max} - 10\ mph)$, then the leader CV needs to slow down to a speed that is lower than $(S_{max} - 10\ mph)$ to reach the intersection before the signal turns green again. However, as we mentioned above, an advised speed lower than $(S_{max} - 10\ mph)$ may seem too low considering the speed of other vehicles on the roadway and the drivers may not want to or able to follow that. In that case, the leader CV is advised a speed equal to $(S_{max} - 10\ mph)$, as this will be the only solution that meets the constraint in (14). On the other hand, if $\frac{d_{L,int}(k)}{t_{avail}(k)} \geq (S_{max} - 10\ mph)$, then we set the minimum value between $S_{max}$ and $\frac{d_{L,int}(k)}{t_{avail}(k)}$ as the upper bound, which leads the optimization to pick a solution that would minimize the delay defined in (13) by allowing the leader CV to operate at a speed within the speed limit so that it can arrive at the intersection when it would turn green again. Thus, the constraint defined in (14) helps to find speed advisory solutions for the leader CVs that would minimize the stopped delay by slowing the CVs down. On the other hand, the objective function defined in (13) pushes the advisory speed solutions toward the speed limit, $S_{max}$ (note that, $S_{L,adv} = S_{max}$ yields $delay_L(k) = 0$ in (13)) and the optimization determines $S_{L,adv}$ that is optimum in terms of the above two opposing conditions.

As mentioned before, the algorithm described in this subsection runs in the CV platoon assigner, a serverless process, i.e., AWS Lambda, as shown in Fig. 2. Once the CV Platoon Assigner assigns the CVs into platoons and determines the advisory speeds for the corresponding leader CVs, it saves the results into the Speed Advisory Database. Then, it invokes CV platoon optimizers, i.e., one CV platoon optimizer for one CV platoon, to run another algorithm of speed advisory optimization for the follower CVs in the platoons, which we explain in the following subsection.

### 3.2.3 Speed Advisory for the Follower CVs in the Platoons

While the leader CVs of the platoons are advised speeds to help the CVs quickly pass the intersection (for case I) or to reduce the stopped delay as much as possible (for case II), the follower CVs are advised speeds simply to reduce the gap among the follower CVs as much as possible without causing any safety issues, such as increased collision risks compared to the case when the CVs run without any advisory speeds. We did this using a discrete-time linear model predictive control (MPC)-based optimization algorithm that is solved globally to determine the speed advisories for all the follower CVs in each platoon at each time step. In this subsection, we will discuss the detailed formulation of the MPC-based optimization for the follower CVs' speed advisories. Fig. 3 and Table 1 present the relevant symbols that are used in this formulation.

First, we assume the advised speeds are achievable by the follower CVs in a platoon within a short period of time $\Delta t$ based on the CVs' maximum acceleration, $a_{Acc}$, or deceleration, $a_{Brk}$, capabilities. Then, assuming constant acceleration or deceleration within this short period of time, $\Delta t$, we can write the following equations of motion for the $i^{th}$ and the $(i+1)^{th}$ follower CVs in a platoon,

$$x_i(k+1) = x_i(k) + \left(\frac{S_i(k)+S_{i,adv}(k)}{2}\right)\Delta t = x_i(k) + u_i(k)\Delta t \tag{15}$$

$$\text{where, } u_i(k) = \left(\frac{S_i(k)+S_{i,adv}(k)}{2}\right) \tag{16}$$

$$\text{similarly, } x_{i+1}(k+1) = x_{i+1}(k) + u_{i+1}(k)\Delta t$$

Now, we estimate the gap ($g_{i+1}(k+1)$) for the $(i+1)^{th}$ follower CV with its immediate leading follower CV, i.e., the $i^{th}$ follower CV, as,

$$g_{i+1}(k+1) = x_{i+1}(k+1) - x_i(k+1) - l_{CV} = g_{i+1}(k) + [u_{i+1}(k) - u_i(k)]\Delta t \tag{17}$$

In this algorithm, we assume that the lengths of all the CVs are the same, i.e., $l_{CV}$ is the same for all the CVs. However, individual CV length can be used as well if the information is available. Note that, (16) stands for the control input that we seek from our MPC-based optimization. Once we obtain the control inputs, we can easily determine the speed advisories for the follower CVs from (16). Now, as (17) is applicable for all the follower CVs in a platoon, we can write it in an augmented matrix form as follows,

$$\boldsymbol{G}(k+1) = \boldsymbol{G}(k) + \boldsymbol{B}\boldsymbol{U}(k) \tag{18}$$

where, $G(k) = \begin{bmatrix} g_L(k) \\ g_1(k) \\ g_2(k) \\ ... \\ g_N(k) \end{bmatrix}$, $B = \begin{bmatrix} 0 & 0 & 0 & \cdots & 0 \\ -\Delta t & \Delta t & 0 & \cdots & 0 \\ 0 & -\Delta t & \Delta t & & 0 \\ & \vdots & & \ddots & \vdots \\ 0 & 0 & 0 & \cdots & \Delta t \end{bmatrix}_{(N+1) \times (N+1)}$, and $U(k) = \begin{bmatrix} u_L(k) \\ u_1(k) \\ u_2(k) \\ ... \\ u_N(k) \end{bmatrix}$

Note that, although the speed advisory for the leader CV in a platoon is not sought from this MPC-based optimization, we still include the leader CV in (18) because the gap associated with $1^{st}$ follower CV in a platoon is calculated with respect to the leader CV of that platoon. However, as the leader CV does not have an immediate leading CV, the dynamics of its gap cannot be formulated as in (17). Therefore, all the entries of the first row of $B$ are set to zeros and $g_L(k)$ is set to an arbitrary value. Thus, the gap for the leader CV, $g_L(k)$, will remain unchanged over the prediction horizon irrespective of whatever control inputs are chosen and it will not affect our MPC-based optimization.

To determine the follower CVs' target gap at each timestamp, we adopted the constant time gap (CTG) policy. In a CTG policy, all the follower CVs in a platoon are expected to maintain a constant time gap with their immediate leading CVs. Besides, in a platooning operation, CTG policy can help reduce the collision risks by varying the target gap requirement based on the speed of the vehicles. In this study, we considered a two-second constant time gap, i.e., $T_g = 2$ seconds, with a two-meter standstill gap, i.e., $g_{stand} = 2$ meters (Mahmod et al., 2013; Park and Schneeberger, 2003). A standstill gap is a minimum gap to avoid the chance of collisions that all CVs must maintain, even if they come to a complete stop. Therefore, the target gap for the $(i+1)^{th}$ follower CVs in a platoon (denoted as $g_{(i+1),tar}(k+1)$) is given by,

$$g_{(i+1),tar}(k+1) = S_{i+1}(k) \times T_g + g_{stand} \quad (19)$$

where, $g_{stand}$ denotes constant standstill distance. As (19) can be written for all the follower CVs in a platoon, we can write them in an augmented form as follows,

$$G_{tar}(k+1) = G_{tar}(k) \quad (20)$$

where, $G_{tar}(k+1) = \begin{bmatrix} g_{L,tar}(k+1) \\ g_{1,tar}(k+1) \\ g_{2,tar}(k+1) \\ ... \\ g_{N,tar}(k+1) \end{bmatrix}$, and $G_{tar}(k) = \begin{bmatrix} S_L(k) \times T_g + g_{stand} \\ S_1(k) \times T_g + g_{stand} \\ S_2(k) \times T_g + g_{stand} \\ ... \\ S_N(k) \times T_g + g_{stand} \end{bmatrix}$

Now, we augment (18) and (20) to get the state dynamics for our MPC-based optimization,

$$\begin{bmatrix} G(k+1) \\ G_{tar}(k+1) \end{bmatrix} = \begin{bmatrix} G(k) \\ G_{tar}(k) \end{bmatrix} + \begin{bmatrix} B \\ 0_{(N+1) \times (N+1)} \end{bmatrix} U(k) \quad (21)$$

where, $0_{(N+1) \times (N+1)}$ is an $(N+1) \times (N+1)$ dimensional matrix with all zero entries. We can rewrite (21) as,

$$X_a(k+1) = A_a X_a(k) + B_a U(k) \quad (22)$$

where, $X_a(k+1) = \begin{bmatrix} G(k+1) \\ G_{tar}(k+1) \end{bmatrix}$, $X_a(k) = \begin{bmatrix} G(k) \\ G_{tar}(k) \end{bmatrix}$, $A_a = I_{2(N+1) \times 2(N+1)}$, and

$$B_a = \begin{bmatrix} B \\ 0_{(N+1) \times (N+1)} \end{bmatrix}$$

where, $I_{2(N+1) \times 2(N+1)}$ is an $(N+1) \times (N+1)$ dimensional identity matrix. As with this MPC-based optimization, we want to adjust the gap among the follower CVs in a platoon based on the CTG policy, we define our measured variable matrix (denoted as $Y_a(k)$) as follows,

$$Y_a(k) = G(k) - G_{tar}(k) = [I_{(N+1) \times (N+1)} \quad -I_{(N+1) \times (N+1)}] \begin{bmatrix} G(k) \\ G_{tar}(k) \end{bmatrix} \quad (23)$$

which can be rewritten as,

$$Y_a(k) = C_a X_a(k) \quad (24)$$

where, $C_a = [I_{(N+1) \times (N+1)} \quad -I_{(N+1) \times (N+1)}]$

Now, we define our cost function for the optimization. In this case, we preferred a quadratic cost function as our aim is to minimize the difference between the current gap, $G(k)$, and the target gap, $G_{tar}(k)$, based on the CTG policy through the speed advisories. Therefore, the cost function for a single-step prediction horizon (as only one step is required to be predicted based on the state dynamics defined in (22)) can be written as,

$$J = Y_a^T(k) Y_a(k) \quad (25)$$

Substituting $Y_a(k)$ from (24) into (25), we get,

$$J = X_a^T(k) C_a^T C_a X_a(k) = X_a^T(k) P X_a(k) \quad (26)$$

where, $P = C_a^T C_a$

Now, we move on to the constraints for this MPC-based optimization. In this case, we introduce constraints for the control inputs defined in (16) and the measured variables defined in (24). First, the follower CVs should never be advised with speeds that exceed the roadway speed limit, $S_{max}$, nor should they be advised negative speeds, which leads us to the following constraint,

$$0 \leq S_{i,adv}(k) \leq S_{max} \quad (27)$$

As each control input is defined as the average of each follower CV's current speed, $S_i(k)$, and advised speed, $S_{i,adv}(k)$, in (16), we can rewrite (27) in terms of the control input as follows,

$$\frac{S_i(k)}{2} \leq u_i(k) \leq \frac{1}{2}(S_i(k) + S_{max}) \quad (28)$$

Second, as mentioned before, we assumed that the advised speeds are achievable by the follower CVs based on their maximum acceleration, $a_{Acc}$, or braking deceleration, $a_{Brk}$, capabilities. Therefore, we also have,

$$S_i(k) + a_{Brk} \Delta t \leq S_{i,adv}(k) \leq S_i(k) + a_{Acc} \Delta t \quad (29)$$

Again, we can rewrite (29) in terms of the control input for the $i^{th}$ follower CV as,

$$S_i(k) + \frac{1}{2} a_{Brk} \Delta t \leq u_i(k) \leq S_i(k) + \frac{1}{2} a_{Acc} \Delta t \quad (30)$$

Then, combining (28) and (30) to get a single equation of constraint for the control input of the $i^{th}$ follower CV,

$$\max\left(\frac{S_i(k)}{2}, \left(S_i(k) + \frac{1}{2} a_{Brk} \Delta t\right)\right) \leq u_i(k) \leq \min\left(\frac{1}{2}(S_i(k) + S_{max}), \left(S_i(k) + \frac{1}{2} a_{Acc} \Delta t\right)\right) \quad (31)$$

We can write (31) into an augmented form as,

$$U_{low}(k) \leq U(k) \leq U_{high}(k) \quad (32)$$

$$\text{where, } \boldsymbol{U_{low}}(k) = \begin{bmatrix} \max\left(\frac{S_L(k)}{2}, \left(S_L(k) + \frac{1}{2}a_{Brk}\Delta t\right)\right) \\ \max\left(\frac{S_1(k)}{2}, \left(S_1(k) + \frac{1}{2}a_{Brk}\Delta t\right)\right) \\ \max\left(\frac{S_2(k)}{2}, \left(S_2(k) + \frac{1}{2}a_{Brk}\Delta t\right)\right) \\ \ldots \\ \max\left(\frac{S_N(k)}{2}, \left(S_N(k) + \frac{1}{2}a_{Brk}\Delta t\right)\right) \end{bmatrix}, \text{ and}$$

$$\boldsymbol{U_{high}}(k) = \begin{bmatrix} \min\left(\frac{1}{2}(S_L(k) + S^{max}), (S_L(k) + \frac{1}{2}a_{Acc}\Delta t)\right) \\ \min\left(\frac{1}{2}(S_1(k) + S^{max}), (S_1(k) + \frac{1}{2}a_{Acc}\Delta t)\right) \\ \min\left(\frac{1}{2}(S_2(k) + S^{max}), (S_2(k) + \frac{1}{2}a_{Acc}\Delta t)\right) \\ \ldots \\ \min\left(\frac{1}{2}(S_N(k) + S^{max}), (S_N(k) + \frac{1}{2}a_{Acc}\Delta t)\right) \end{bmatrix}$$

Next, we introduce a lower bound for the measured variable, $\boldsymbol{Y_a}(k)$, due to safety considerations. As the optimized solution should not result in a situation where any of the follower CVs has a lower gap than its corresponding target gap based on the CTG policy, we write,

$$\boldsymbol{Y_a}(k) \geq \boldsymbol{0}_{(N+1)\times 1} \tag{33}$$

Now, we have all the necessary equations formulated that we need for our MPC-based speed advisory optimization for the follower CVs in a platoon. As our linear MPC formulation includes a quadratic cost function (as given in (26)), we utilized a Python-based open-source solver, i.e., CVXOPT ("CVXOPT," 2022), for solving quadratic programming problems to run this MPC-based optimization. As mentioned before, this part of the algorithm runs in the CV Platoon Optimizer as shown in Fig. 2.

## 4. EVALUATION OF CLOUD-BASED CV SPEED ADVISORY APPLICATION

In this section, we present the details of the evaluation method (i.e., cloud-in-the-loop simulation), evaluation metrics, results, and discussions for our serverless cloud-based CV speed advisory application.

### 4.1. Cloud-in-the-loop Simulation

We conducted three case studies for different roadway traffic conditions by developing a cloud-in-the-loop simulation testbed to evaluate the feasibility of the developed serverless CV speed advisory application at a system level. In addition, we compared the results obtained from the simulation with and without the application to evaluate the efficacy of the application in terms of stopped delay of the CVs at the signalized intersections, total travel time of the CVs through the signalized roadway section, and an aggregated collision risk indicator. Then, we evaluated the communication and processing delays for running the serverless CV speed advisory application to evaluate the feasibility of such CV-based roadway traffic management application in terms of latency requirement.

We used an open-source microscopic multimodal traffic simulator called Simulation Urban Mobility (SUMO) developed by the German Aerospace Center (Lopez et al., 2018). We used SUMO to simulate a section of a roadway including traffic signals and CVs operating in the roadway section. In our cloud-in-the-loop simulation, AWS services (residing in the cloud), such as DynamoDB and KDS, are integrated with SUMO (running in a local machine) to evaluate the serverless CV speed advisory application (as shown in Fig. 4). Traffic Control Interface (TraCI) (Wegener et al., 2008) is a Python-based interface compatible with SUMO. As Fig. 4 shows, we used TraCI to extract BSMs (e.g., CVs' location and motion information) from CVs and signal phase and timing messages (e.g., current signal interval, remaining green time) from the traffic signals simulated in SUMO. Data collected from the simulation are forwarded to the AWS cloud through different AWS services, i.e., DynamoDB and KDS, via Long Term Evolution (LTE) communication. As shown in Fig. 4, the CV Trajectory Database (based on DynamoDB) receives and stores the most recent CV trajectory information from SUMO via TraCI. In the cloud, each KDS triggers a Serverless CV Advisory Cluster, as mentioned before. Each Serverless CV Advisory Cluster gets CV trajectory information from the CV Trajectory Database. Each Serverless CV Advisory Cluster also collects and updates the distances of the CVs from its corresponding traffic signal (as shown in Fig. 4). As shown in Fig. 2, each Serverless CV Advisory Cluster consists of several Serverless CV Advisory Modules. Each of these modules include a CV Platoon Assigner (based on a Lambda function) and a number of CV Platoon Optimizers (also based on Lambda functions) equal to the number of platoons assigned by the corresponding CV Platoon Assigner. Inside each Serverless CV Advisory Module, the CV platoon identification algorithm runs on a CV Platoon Assigner function, and the speed advisory optimization algorithms run on CV Platoon Optimizers (details are mentioned in the previous section). The results of the optimizations, i.e., the speed advisories are saved in the Speed Advisory Database (also based on DynamoDB). Then, SUMO, running in our local machine, can collect the speed advisories via LTE and assign the speed advisories to the CVs through TraCI.

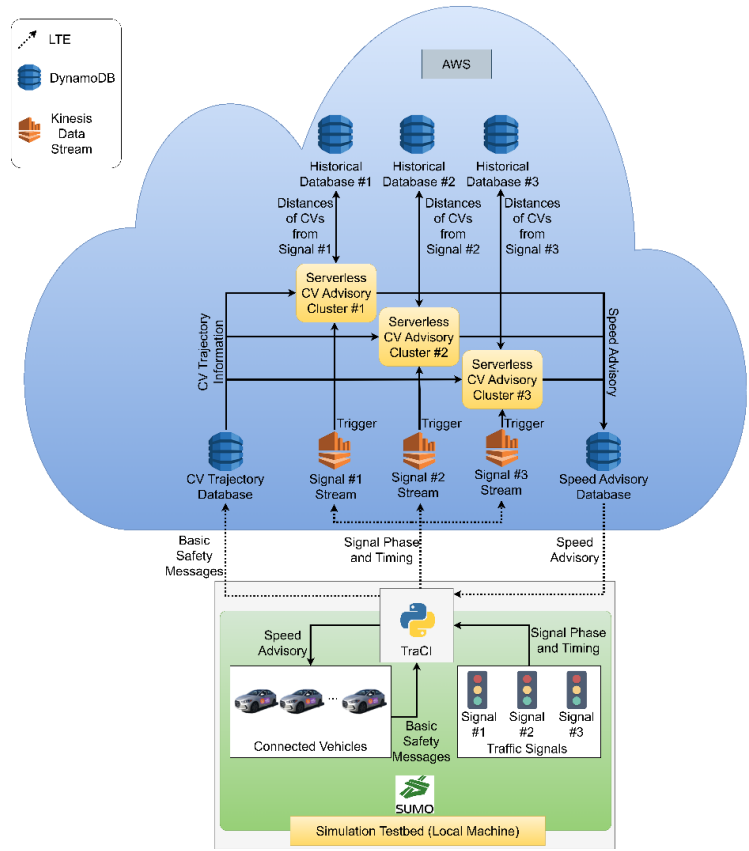

**Fig. 4.** Dataflow in the cloud-in-the-loop simulation.

In Fig. 5, the simulated roadway is shown in orange-coloured line, which is a 1.5-mile-long 4-lane highway (2 lanes in each direction) with three traffic signals in Clemson, South Carolina, and it is a part of a CV deployment site known as South Carolina Connected Vehicle Testbed (SC-CVT) (Chowdhury et al., 2018). Since this testbed hosts a suite of CV deployment-related technologies, such as roadside computing and communication infrastructures, and cellular towers, choosing this site would help us validate, update, and expand the CV application developed in this study through rigorous field experiments in future. By defining the traffic flows in SUMO configuration ("Shortest or Optimal Path Routing - SUMO Documentation," 2022), we generated 50 CVs on the simulated roadway in three different traffic densities, i.e., low, medium, and high traffic densities. SUMO allows controlling the time interval within which a given number of vehicles will be generated, which we used here to create the different traffic densities. Here, low traffic density stands for 633 passenger cars per hour per lane (pc/h/ln), which is 33% of the traffic capacity, i.e., 1900 pc/h/ln, medium traffic density stands for 1267 pc/h/ln, i.e., 66% of traffic capacity, and high traffic density stands for 1900 pc/h/ln, i.e., full traffic capacity, based on the base saturation flowrate defined in ("Highway Capacity Manual 2010," 2010). All CVs operate within a roadway speed limit of 35 mph, which is already included in the map data. For each condition, we evaluated two scenarios in the simulation: 1) the baseline scenario, i.e., no speed advisory, and 2) the serverless CV speed advisory application-deployed scenario. For each traffic density defined above, we ran the simulation multiple times with randomly generated CVs.

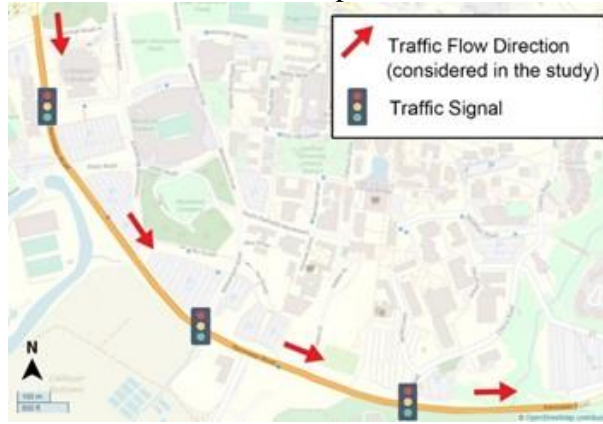

**Fig. 5.** Route location and layout.

### 4. 2. Evaluation Metrics

We evaluated the end-to-end delay to assess the feasibility of the serverless CV speed advisory application as a real-time CV mobility application. The end-to-end delay is calculated using the following equation,

$$\text{end-to-end delay} = \text{upload delay} + \text{processing time} + \text{download delay} \qquad (34)$$

Here, the upload delay for each CV is calculated as the delay observed from a CV to upload its trajectory information into the CV Trajectory Database. The processing time for each Serverless CV Advisory Cluster is calculated as the time taken by that cluster from being triggered by its corresponding traffic signal to generate the CV speed advisories and save them into the Speed Advisory Database. The download delay for each CV is the delay observed between a download request submitted by that CV to the cloud and receival of the speed advisory.

To evaluate whether our serverless CV speed advisory application is effective as a CV-based roadway traffic management application in terms of roadway traffic condition and safety, we compared three measures of effectiveness (MoEs): 1) stopped delay at the signalized intersections of the simulated roadway, 2) total travel time to pass the simulated roadway section, and 3) time-integrated time-to-collision (TIT). Stopped delay and travel time are MoEs related to traffic flow, whereas TIT is a widely used surrogate measure for evaluating collision risks that integrates the time-to-collision (TTC) profile below a predefined threshold (i.e., TTC threshold, $TTC^*$) over time for all the CVs under collision risk evaluation (Wu et al., 2021; Yang et al., 2017). Details for calculating TTC and TIT can be found in (Minderhoud and Bovy, 2001; Shi et al., 2018). For our study, TIT for the $i^{th}$ CV (i.e., $TIT_i$) can be calculated using the following equation,

$$TIT_i = \sum_t [TTC^* - TTC_i(t)], \quad \forall 0 \leq TTC_i(t) \leq TTC^* \tag{35}$$

$$\text{where, } TTC_i(t) = \begin{cases} \frac{g_i(t)}{S_i(t) - S_{i-1}(t)} & if\ S_i(t) > S_{i-1}(t) \\ \infty & if\ S_i(t) \leq S_{i-1}(t) \end{cases} \tag{36}$$

Here, $t$ represents a timestamp, $g_i(t)$ represents the gap between the $i^{th}$ CV (i.e., a follower CV) and the $(i-1)^{th}$ CV (i.e., immediate leading CV of the $i^{th}$ CV) at $t$, and $S_i(t)$ and $S_{i-1}(t)$ represent the speeds of the $i^{th}$ and the $(i-1)^{th}$ CVs at $t$, respectively. As observed from (35) and (36), the risk of collision is only considered when the follower CV has a higher speed compared to its immediate leading CV. Once TIT for all the CVs is calculated using (35), we sum them up to get the aggregated TIT for all the CVs within the simulation run time. In this paper, we used a $TTC^*$ of 2 seconds based on the time headway requirement in our MPC-based optimization for determining the advisory speeds of the follower CVs. A $TTC^*$ of 2 seconds implies a risk of collision whenever the time gap between any two successive CVs is less than or equal to 2 seconds.

### 4. 3.    Evaluation Results and Discussions

Fig. 6 presents the processing time, and the end-to-end delay found from our experiments and Table 2 shows the average processing time in the cloud and the average end-to-end delay of the application for each CV under the three traffic density conditions considered in this study. Fig. 6 (a) shows that the average processing delay in the cloud varies within 5 ms for the three traffic density conditions, and Fig. 6 (b) shows that the average end-to-end delay of the application for each CV varies within 20 ms. This implies that the change in roadway traffic conditions does not affect the processing time and the end-to-end latency, which proves the scalability of the serverless cloud-based CV application developed in this study. From Table 2, the end-to-end delay is about 452 ms (on average for all three traffic density conditions), which meets the requirement of a real-time CV mobility application, i.e., maximum allowable delay of 1000 ms (Fehr, 2014; Islam et al., 2020). Note that, the results presented in Fig. 6 and Table 2 for our serverless CV speed advisory indicates the feasibility of implementing such real-time CV-based roadway traffic management applications using serverless public cloud infrastructure in terms of scalability and latency requirement.

Fig. 7 shows box chart comparisons between our serverless CV speed advisory application and the baseline "no speed advisory" scenario in terms of stopped delay and total travel time for three different roadway traffic conditions, i.e., low, medium, and high-density traffic. As observed from Fig. 7(a), our serverless CV speed advisory application reduced the stopped delay for all three roadway traffic conditions compared to the baseline "no speed advisory" scenario. The speed advisory algorithm developed in this study aims to advise the CVs to operate at the maximum allowable speed (if the algorithm estimates that the traffic signal will display green when the CVs reach the intersection), or to slow down to an extent so that the CVs can get the green light they reach the intersection (if it is not green already). Thus, the reduction in the stopped delay observed in Fig. 7(a) is expected. In Fig. 7(b), we observe a small reduction in the total travel time when using the serverless CV speed advisory application for providing speed advisories to the CVs as compared to the "no speed advisory" case. This is not unexpected because our speed advisory optimization aims to reduce the stopped delay, not the travel time. While it may seem that a reduction in the stopped delay should cause a reduction in the travel time as well, it may not be the case all the time (Eckhoff et al., 2013). For example, note that although the serverless CV speed advisory application reduces the total stopped delay for the CVs significantly, it cannot entirely remove the stopped delay and the CVs may have to stop at the intersections for some time. Then, these CVs would have to start from a stopped condition when the signal turns green again in which case the benefit of having no startup lost time is not achievable. Also, our serverless CV speed advisory application does not advise CVs with speeds considering that they can pass the intersection within the yellow interval. On the other hand, in the "no speed advisory" case, the CVs have no such conditions imposed on them. Thus, reducing the total travel time is not always guaranteed for all the CVs while using this serverless CV speed advisory application.

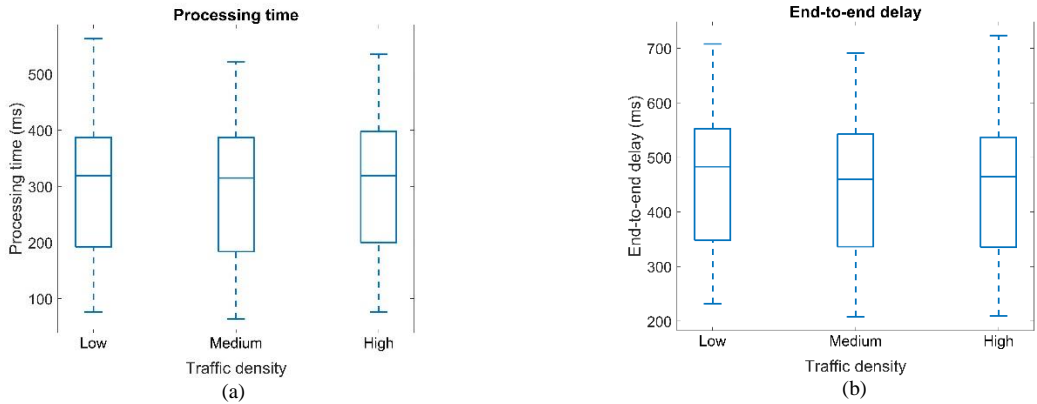

**Fig. 6.** Box charts of (a) processing time, and (b) end-to-end delay.

**Table 2**
Average processing time and end-to-end delay for the serverless CV speed advisory application.

|  | Traffic density | | | Average of the three traffic densities | Allowable delay |
|---|---|---|---|---|---|
|  | Low | Medium | High | | |
| Average processing time | 298 | 297 | 303 | 299 | |
| Average end-to-end delay | 463 | 447 | 446 | 452 | <1000 ms |

Table 3 presents the effectiveness of the serverless CV speed advisory application in terms of percentage reduction of the selected MoEs on average for each CV in the simulation. As shown in Table 3, our application was able to reduce the average (per CV) stopped delay by 77%, the average (per CV) total travel time by 3%, and the average (per CV) TIT by 21% for the three roadway traffic density conditions considered in this study. We observe that the maximum reduction of the stopped delay, i.e., about 85%, was possible for low traffic density. In terms of reducing the total travel time, our serverless CV speed advisory application's performance did not vary much based on the different traffic conditions. We also observe that the application is most effective in reducing the average per CV TIT, i.e., about 24%, for low-density traffic condition.

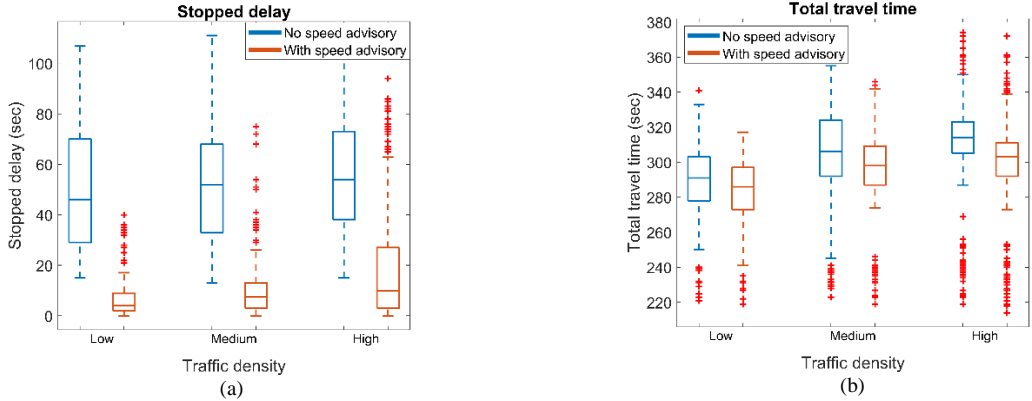

**Fig. 7.** Box chart comparisons for (a) stopped delay and (b) total travel time.

**Table 3**
Average (per CV) reduction of the MoEs for the serverless CV speed advisory application.

| Average percentage | Roadway traffic density | | | Average of low, medium, and |
| reduction in | Low | Medium | High | high traffic densities |
| --- | --- | --- | --- | --- |
| Stopped delay | 85% | 80% | 65% | 77% |
| Total travel time | 2% | 3% | 4% | 3% |
| TIT | 24% | 16% | 23% | 21% |

## 5. CONCLUSIONS

In this paper, we presented an automated and scalable serverless public cloud-based real-time CV speed advisory application while meeting the strict latency threshold of CV mobility applications. The serverless CV speed advisory application assists CVs to pass through a signalized corridor with speed advisories that can help reduce the stopped delay experienced by the CVs at the intersections.

We presented a serverless public cloud-based architecture to support CV mobility applications that require the cloud to communicate with the CVs and the transportation infrastructures in a roadway. Later, we fine-tuned the architecture to support our CV speed advisory application in an automated and scalable fashion. Our optimization-based real-time CV speed advisory algorithm was developed following a modular design approach. First, an algorithm identifies CV platoons based on whether they can pass a signalized intersection within the available time of the current green time or the next green time. Second, once the CV platoons are identified, another algorithm determines optimal speed advisories for the leader CVs of the respective platoons. These advisories provided to the CVs are optimized to minimize the stopped delay at the intersections, i.e., the algorithm determines optimal speed advisories for the leader CVs so that they can either speed up to pass the intersection within the current green time or slow down to an extent that would minimize their stopped delay at the intersections while waiting for the next green time. Finally, a third algorithm determines the optimal speed advisories for the follower CVs in the platoons to minimize the inter-vehicle gaps while considering minimum safety gaps using an MPC-based approach. The first two algorithms are run using an AWS Lambda, which invokes a suite of AWS Lambdas to run in parallel to determine the optimal speed advisories for the follower CVs of the platoons using the third algorithm, i.e., an optimal speed advisory algorithm for the follower CVs in a platoon is run using a dedicated AWS Lambda. Leveraging the parallel computing technique here ensures the scalability of the application, i.e., the application can immediately scale up (i.e., utilize additional resources in the cloud) or down (i.e., utilizes reduced resources in the cloud) based on the roadway traffic densities.

We conducted experiments on a simulated signalized corridor for three different roadway traffic conditions (low, medium, and high-density roadway traffic) with a cloud-in-the-loop simulation testbed using AWS and SUMO to evaluate the feasibility and performance of the serverless CV speed advisory application at a system level. Our experiments indicated that our serverless CV speed advisory application reduced the average stopped delay at the intersections by 77%, the average travel time through a signalized corridor by 3%, and the average time-integrated time-to-collision by 21%, which proved the effectiveness of the application. The average end-to-end delay of our application was reported 452 ms, which is well under the maximum allowable end-to-end latency for CV mobility applications' latency of 1000-ms. This proved the feasibility of our application to be deployed as a real-time CV speed advisory application.

Based on the evaluation results, we conclude that the serverless cloud architectures built upon public cloud infrastructure are capable of providing promising solutions to implementing real-time CV-based roadway traffic management applications. Indeed, public cloud infrastructure can be considered by public and private transportation agencies to implement real-time CV-based roadway traffic management applications as it would alleviate the need for significant investment in upgrading the legacy transportation infrastructure to support such applications.

## 6. LIMITATIONS AND FUTURE SCOPES

In this study, we developed a CV mobility application that uses public cloud services, which was able to meet the corresponding latency requirements. However, the results presented in this study are obtained under certain assumptions associated with the simulations we conducted, such as the number of lanes, lane widths, roadway speed limits, shared or dedicated left/right turn lanes, and CV penetration levels. Further studies (both simulations and field experiments) are required to investigate the effects of these parameters on the application's performance.

We utilized services offered by the AWS only to develop our serverless CV speed advisory application. Our future studies will explore the other public cloud services, such as the Google Cloud Platform (GCP), Microsoft Azure, IBM Cloud, and Oracle Cloud for developing similar applications since some services offered by certain cloud service providers might be more useful for certain applications than those of the other cloud service providers.

This study did not evaluate how beneficial it is to utilize public cloud services for CV applications compared to using roadside computing infrastructure. Further studies are needed to investigate the advantages and disadvantages of utilizing public cloud services over building disruptive roadside computing infrastructure to support real-time CV applications.

## ACKNOWLEDGMENTS AND DECLARATIONS

This work is based upon the work partially supported by the Center for Connected Multimodal Mobility ($C^2M^2$) (a U.S. Department of Transportation Tier 1 University Transportation Center) headquartered at Clemson University, Clemson, SC, USA. Any opinions, findings, conclusions, and recommendations expressed in this material are those of the author(s) and do not necessarily reflect the views of $C^2M^2$, and the US Government assumes no liability for the contents or use thereof.

The authors declare that the contents of this article have not been published previously. The authors confirm contribution to the paper as follows: study conception and design: M.S. Salek,

M. Rahman, M. Chowdhury; analysis and interpretation of results: H.-W. Deng, M.S. Salek, M. Rahman; manuscript preparation: H.-W. Deng, M.S. Salek, M. Rahman, M. Chowdhury, M. Shue, A.W. Apon. All authors reviewed the manuscript and approved the contents for publication in this journal. All the authors have no conflict of interest with the funding entity and any organization mentioned in this article that may have influenced the conduct of this research and the findings. The authors declare that this manuscript did not require any human subject research.